\documentclass[journal]{IEEEtran}

\usepackage{times}
\usepackage{multirow}
\usepackage{xcolor}
\usepackage{booktabs}
\usepackage{hyperref}
\usepackage{tikz}

\usepackage{graphicx}
\ifCLASSINFOpdf
\else
\fi
\usepackage{amsmath}
\usepackage{amssymb}
\usepackage{amsthm}
\usepackage{mathtools}
\usepackage{diagbox}
\usepackage[ruled, lined,noend, linesnumbered, commentsnumbered]{algorithm2e}
\graphicspath{{Figs/}}

\DeclareMathOperator*{\argmin}{arg\,min}

\newtheoremstyle{definition}
{\topsep}{\topsep}%
{\itshape}{}%
{\bfseries}{}%
{\newline}{}%
\theoremstyle{definition}
\newtheorem{definition}{Definition}

\definecolor{commentsColor}{RGB}{219, 48, 122}
\newcommand\mycommfont[1]{\footnotesize\ttfamily\textcolor{commentsColor}{#1}}
\SetCommentSty{mycommfont}
\let\oldnl\nl
\newcommand{\nonl}{\renewcommand{\nl}{\let\nl\oldnl}}

\usepackage[normalem]{ulem}
\useunder{\uline}{\ul}{}

\begin{document}
%
\title{A fast and Accurate Similarity-constrained Subspace Clustering Framework for Unsupervised Hyperspectral Image Classification}
%
%

\author{Carlos~Hinojosa,~\IEEEmembership{Student~Member,~IEEE,}
        Esteban~Vera,~\IEEEmembership{Member,~IEEE,}
        and~Henry~Arguello,~\IEEEmembership{Senior~Member,~IEEE}
\thanks{Carlos Hinojosa and Henry Arguello are with the Department of System Engineering and Informatics, Universidad Industrial de Santander, Bucaramanga, Colombia (e-mail: carlos.hinojosa@saber.uis.edu.co; henarfu@uis.edu.co).}
\thanks{Esteban Vera is with the School of Electrical Engineering, Pontificia Universidad Cat\'olica de Valpara\'iso, Valpara\'iso, Chile (e-mail: esteban.vera@pucv.cl).}
\thanks{Manuscript received April 19, 2005; revised August 26, 2015.}}

%
%

\markboth{}%
{}
%



\maketitle

\begin{abstract}
Accurate land cover segmentation of spectral images is challenging and has drawn widespread attention in remote sensing due to its inherent complexity. Although significant efforts have been made for developing a variety of methods, most of them rely on supervised strategies. Subspace clustering methods, such as Sparse Subspace Clustering (SSC), have become a popular tool for unsupervised learning due to their high performance. However, the computational complexity of SSC methods prevents their use on large spectral remotely sensed datasets. Furthermore, since SSC ignores the spatial information in the spectral images, its discrimination capability is limited, hampering the clustering results' spatial homogeneity. To address these two relevant issues, in this paper, we propose a fast algorithm that obtains a sparse representation coefficient matrix by first selecting a small set of pixels that best represent their neighborhood. Then, it performs spatial filtering to enforce the connectivity of neighboring pixels and uses fast spectral clustering to get the final segmentation. Extensive simulations with our method demonstrate its effectiveness in land cover segmentation, obtaining remarkable high clustering performance compared with state-of-the-art SSC-based algorithms and even novel unsupervised-deep-learning-based methods. Besides, the proposed method is up to three orders of magnitude faster than SSC when clustering more than $2 \times 10^4$ spectral pixels.
\end{abstract}

\begin{IEEEkeywords}
Hyperspectral image classification, Spectral-spatial classification, Subspace clustering, Land-cover segmentation, Unsupervised learning
\end{IEEEkeywords}

%
\IEEEpeerreviewmaketitle

\section{Introduction}
\label{sec:introduction}

\begin{figure}
	\begin{center}
		\includegraphics[width=\columnwidth]{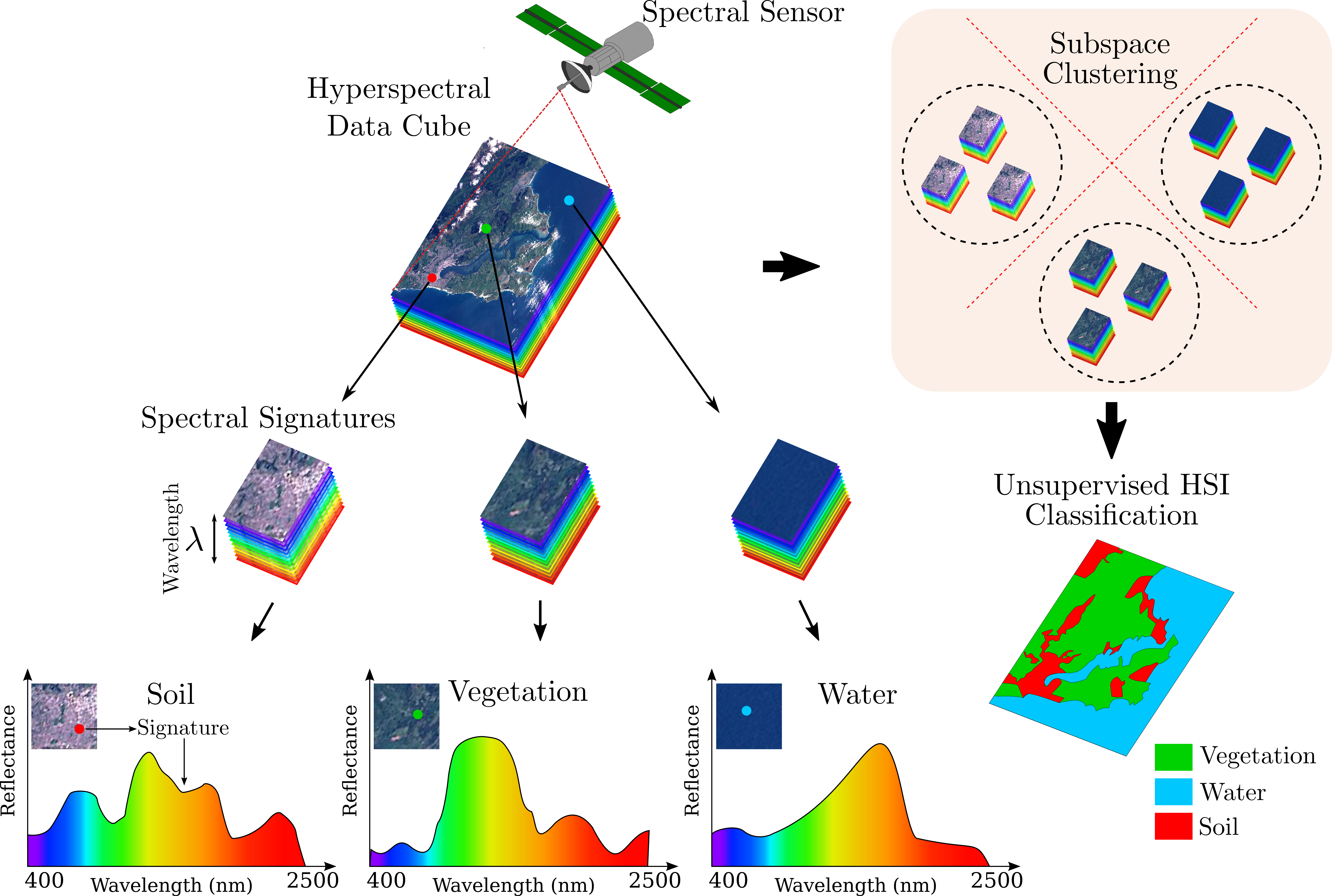}
	\end{center}
	\caption{Unsupervised spectral image land cover segmentation.}
	\label{fig:intro}
\end{figure}

Spectral remote sensing systems acquire information of the Earth's surface by sensing a large amount of spatial data at different electromagnetic radiation frequencies. Spectral images (SI) are commonly regarded as three dimensional datasets or data cubes with two dimensions in the spatial domain $(x,y)$ and one in the spectral domain $(\lambda)$ \cite{shaw2003spectral}. Based on the acquired spectral/spatial resolution, spectral imaging sensors can be categorized in Hyperspectral (HS) and Multispectral (MS). Typically, HS devices capture hundreds of spectral bands of the scene, however, their spatial resolution is often lower compared to that obtained with a MS sensor, which has a low spectral resolution \cite{yokoya2017hyperspectral}.

As shown in Fig. \ref{fig:intro}, every spatial location in a spectral image is represented by a vector whose values correspond to the intensity at different spectral bands. These vectors are also known as the spectral signature of the pixels or spectral pixels. Since different materials usually reflect electromagnetic energy differently at specific wavelengths \cite{shaw2003spectral}, the information provided by the spectral signatures allows distinguishing different physical materials and objects within an image. In remote sensing, the classification of spectral images is also referred to as land cover segmentation or mapping and it is an important computer vision task for many practical applications, such as precision agriculture \cite{lanthier2008hyperspectral}, vegetation classification \cite{thenkabail2016hyperspectral}, monitoring and management of the environment \cite{gessesse2015model,volpi2015semantic}, as well as security and defense issues \cite{briottet2006military}.

Accurate land cover segmentation is challenging due to the high-dimensional feature space and it has drawn widespread attention in remote sensing \cite{ghamisi2017advances,li2019deep}. In the past decade, significant efforts have been made in the development of numerous SI classification methods, however, most of them rely on supervised approaches \cite{sanchez2019supervised,hinojosa2019spectral}. More recently, with the blooming of deep learning techniques for big data analysis, several deep neural networks have been developed to extract high-level features of SIs achieving state-of-the-art supervised classification performance \cite{paoletti2019deep}. However, the success of such deep learning approaches hinges on a large amount of labeled data, which is not always available and often prohibitively expensive to acquire. As a result, the computer vision community is currently focused on developing unsupervised methods that can adapt to new conditions without requiring a massive amount of data.\cite{Kolesnikov_2019_CVPR}.

Most successful unsupervised learning methods exploit the fact that high dimensional datasets can be well approximated by a union of low-dimensional subspaces. Under this assumption, the sparse subspace clustering (SSC) algorithm captures the relationship among all data points by exploiting the \textit{self-expressiveness} property \cite{elhamifar2013sparse}. This property states that each data point in a union of subspaces can be written as a linear combination of other points from its own subspace. Then, the set of solutions is restricted to be sparse by minimizing the $\ell_1$ norm. Finally, an affinity matrix is built using the obtained sparse coefficients, and the normalized spectral clustering algorithm \cite{von2007tutorial} is applied to achieve the final segmentation.

\begin{figure}[t]
	\begin{center}
		\includegraphics[width=\columnwidth]{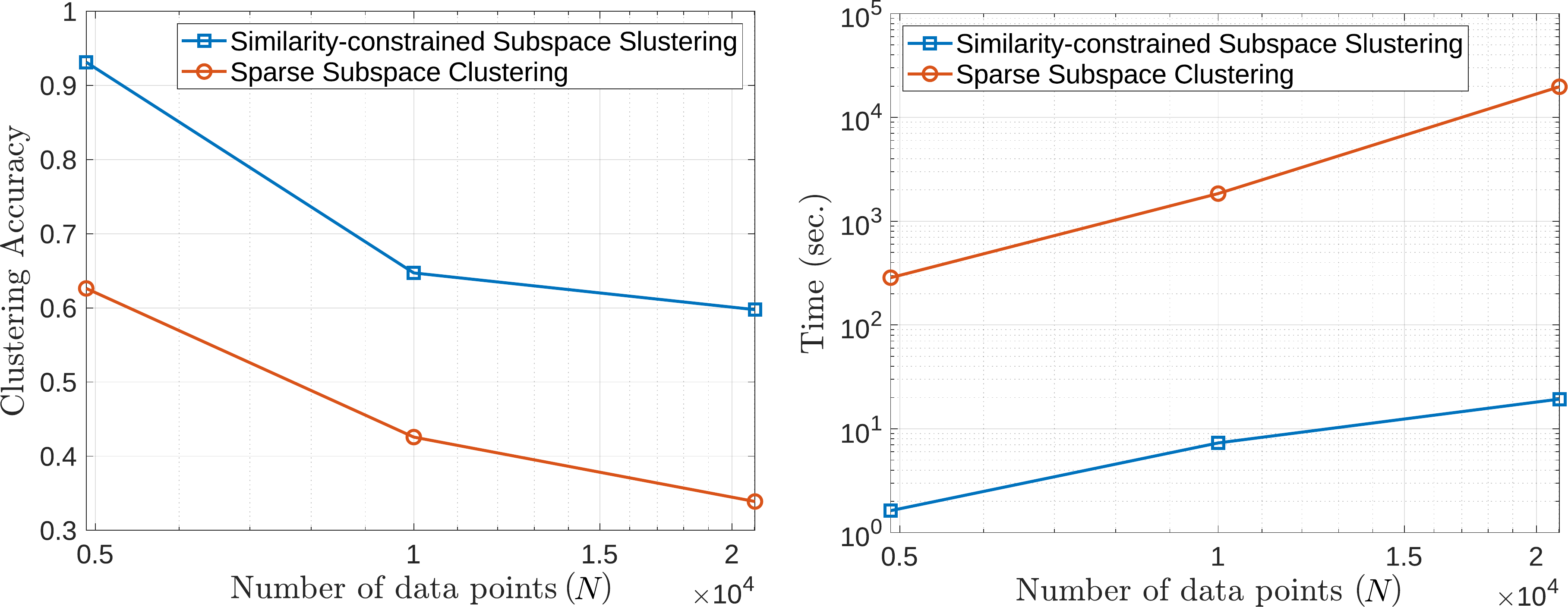}
	\end{center}
	\caption{Clustering accuracy (left) and running time (right) of the SSC algorithm compared with the proposed method for land cover segmentation. In this example, we performed the two subspace clustering algorithms on the full image and two regions of interest (ROIs) of the Indian Pines dataset (See Section \ref{sec:experiments}). The first ROI has $N=4900$ pixels and $k=4$ classes; the second has $N=10000$ pixels and $k=12$ classes; the whole Indian Pines image has $N=21025$ pixels and $k=17$ classes.}
	\label{fig:acc_time_comparison}
\end{figure}

Assuming that spectral pixels with a similar spectrum approximately belong to the same low-dimensional structure, the SSC algorithm can be successfully applied for land cover segmentation. \cite{hinojosa2018coded,hinojosa2018spectral,zhang2016spectral,zhai2016new,huang2019semisupervised}. Despite the great success of SSC in land cover segmentation, two main problems have been identified: (1) The overall computational complexity of SSC prohibits its usage on large spectral remote sensing datasets. For instance, given a SI with $N_r$ rows, $N_c$ columns, and $L$ spectral bands, SSC needs to compute the $N\times N$ sparse coefficient matrix corresponding to $N=N_rN_c$ spectral pixels, whose computational complexity is $O(LN^3)$. Moreover, after building the affinity matrix, spectral clustering performs an eigenvalue decomposition over the $N\times N$ graph Laplacian matrix which also has cubic time complexity, or quadratic using approximation algorithms \cite{chen2018spectral} (see Fig. \ref{fig:acc_time_comparison} right). (2) Under the context of SI, the SSC model only captures the relationship of pixels by analyzing the spectral features without considering the spatial information. Indeed, the sparse coefficient matrix is piecewise smooth since spectral pixels belonging to the same land cover material are arranged in a common region; hence there is a spatial relationship between the representation coefficient vector of one pixel and its neighbors.

\textbf{Paper contribution.} This paper proposes a fast and accurate similarity-constrained subspace clustering algorithm to enhance both the clustering accuracy and execution time when performing land cover segmentation. Specifically, our main contributions are as below


\begin{enumerate}
	\item We propose to first group similar spatial neighboring pixels in subsets using a ``superpixels'' technique \cite{li2015superpixel,achanta2010slic}. Then, instead of expressing each pixel as a linear combination of all pixels in the dataset, we constrain each pixel to be solely represented as a linear combination of other pixels in the same subset. Therefore, the obtained sparse coefficient matrix encodes information about similarities between the most representative pixels of each subset and the whole dataset. In this paper, we present an efficient algorithm for selecting the most representative pixels of each subset by minimizing the maximum representation cost of the data.
	\item Our second contribution is the enhancement of the obtained sparse coefficient matrix via 2D smoothing convolution before applying a fast spectral clustering algorithm that significantly reduces the computational cost. Specifically, the proposed method enforces the connectivity in the affinity matrix and then efficiently obtains spectral embedding without the need to compute the eigenvalue decomposition that has a computational complexity of $O(N^3)$ in general.
\end{enumerate}

Increasing the number of data points and the classes enlarges the computation time and make clustering more challenging. The proposed method, shown with the blue line in the in Fig. \ref{fig:acc_time_comparison}, can be up to three orders of magnitude faster than SSC and outperforms it in terms of accuracy when clustering more than $2 \times 10^4$ spectral pixels. This paper evaluates and compares our approach on three real remote sensing spectral images with different imaging environments and spectral-spatial resolution.

\section{Related Works}
\label{sec:related_works}

In the literature, the scalability issue of SSC and its ability to perform land cover segmentation on spectral images have been studied separately. In this section, we review some related works from these two points of view. Considering a given collection of $N$ data points $\mathbf{X}=\left\{\mathbf{x}_1,\cdots,\mathbf{x}_{N} \right\}$ that lie in the union of $k$ linear subspaces of $\mathbb{R}^D$, SSC expresses each data point $\mathbf{x}_j$ as a linear combination of all other points in $\mathbf{X}$, i.e., $\mathbf{x}_j = \sum_{i \ne j} c_{ij}\mathbf{x}_i$, where $c_{ij}$ is nonzero only if $\mathbf{x}_i$ and $\mathbf{x}_j$ are from the same subspace, for $(i,j) \in \left\{ 1,\cdots, N \right\}$. Such representations $\left\{c_{ij}\right\}$ are called \textit{subspace-preserving}. In general, assuming that $\mathbf{c}_j$ is sparse, SSC solves the following optimization problem
\begin{equation}
	\min_{\mathbf{c}_j \in \mathbb{R}^N} \|\mathbf{c}_j\|_1 + \frac{\tau}{2}\|\mathbf{x}_j - \sum_{i \neq j} c_{ij}\mathbf{x}_i \|_2^2,
	\label{eq:ssc} 
\end{equation}
where $\tau >0$ and $\mathbf{c}_j = \left[ c_{1j},\cdots,c_{Nj} \right]^T$ encodes information about membership of $\mathbf{x}_j$ to the subspaces. Subsequently, an affinity matrix between any pair of points $\mathbf{x}_i$ and $\mathbf{x}_j$ is defined as $A_{ij}=|c_{ij}| + |c_{ji}|$ and it is used in a spectral clustering framework to infer the clustering of the data \cite{von2007tutorial,elhamifar2013sparse}. Although the representation produced by SSC is guaranteed to be subspace preserving, the affinity matrix may lack \textit{connectedness} \cite{nasihatkon2011graph}, i.e., the data points from the same subspace may not form a connected component of the affinity graph due to the sparseness of the connections, which may cause over-segmentation.

\subsection{Fast and Scalable Subspace Clustering Methods}
\label{sec:related_works_sub2}

Taking into account the self-expressiveness property, an early approach to address the SSC scalability issue assumes that a small number of data points can represent the whole dataset without loss of information. Then, authors in \cite{peng2013scalable} proposed the \textit{Scalable Sparse Subspace Clustering} (SSSC) algorithm to cluster a small subset of the original data and then classify the rest of the data based on the learned groups. However this strategy is suboptimal since it sacrifices clustering accuracy for computational efficiency.

In \cite{you2016scalable}, authors replace the $\ell_1$ optimization in the original SSC algorithm \cite{elhamifar2013sparse} with greedy pursuit, e.g., orthogonal matching pursuit (OMP) \cite{tropp2007signal}, for sparse self-representation \cite{dyer2013greedy}. While SSC-OMP improves the time efficiency of SSC by several orders of magnitude, it significantly loses clustering accuracy \cite{chen2017active}. Besides, SSC-OMP also suffers from the connectivity issue presented in the original SSC algorithm. To solve this issue, authors in \cite{you2016oracle} proposed to mixture the $\ell_1$ and $\ell_2$ norms to take advantage of subspace preserving of the $\ell_1$ norm and the dense connectivity of the $\ell_2$ norm. Specifically, this algorithm, named ORacle Guided Elastic Net solver (ORGEN), proposed to identify a support set for each sample. However, in this approach, a convex optimization problem is solved several times for each sample which limits the scalability of the algorithm.

More recent works \cite{aldroubi2018similarity,aldroubi2017cur,abdolali2019scalable} use a different subset selection method for subspace clustering. In particular, the method named Scalable and Robust SSC (SR-SSC) \cite{abdolali2019scalable} selects a few sets of anchor points using a randomized hierarchical clustering method. Then, within each set of anchor points, it solves the LASSO \cite{tibshirani1996regression} problem for each data point, allowing only anchor points to have non-zero weights. However, this method does not demonstrate that their selected points are representative of the subspaces.

Similar to the SSC-OMP paper, authors in \cite{you2018scalable} proposed an approximation algorithm~\cite{williamson2011design} to solve the optimization problem in Eq. (\ref{eq:ssc}). Specifically, instead of using all the dataset $\mathbf{X}$, the Exemplar-based Subspace Clustering (ESC) algorithm in \cite{you2018scalable} selects a small subset $\mathbf{\hat{X}} \subseteq \mathbf{X}$ that represents all data points, and then each point is expressed as a linear combination of points in $\mathbf{\hat{X}} \in \mathbb{R}^{D\times M}$, where $M<N$. In particular, the selection of $\mathbf{\hat{X}}$ is obtained by using the Farthest first search (FFS) algorithm, which is a modified version of the Farthest-First Traversal (FarFT) algorithm~\cite{williamson2011design}. Indeed, the main difference between FarFT and FFS is the used distance metric. Explicitly, FarFT uses the Euclidean distance, while FFS uses a custom metric, derived from Eq. \ref{eq:ssc}, that geometrically measures how well a data point $\mathbf{x}_j \in \mathbf{X}$ is covered by a subset $\mathbf{\hat{X}}$. The authors propose to construct $\mathbf{\hat{X}}$ by first performing random sampling to select a base point and then progressively add new representative data points using the defined metric. However, a careful selection of the search space and the first selected data point could speed up the unsupervised learning process. The complete algorithm proposed in \cite{you2018scalable} is known as ESC-FFS, and we compare it against our proposed method in Section \ref{sec:experiments}.

In general, the previously described algorithms provide an acceptable subspace clustering performance on large-scale datasets. However, these general-purpose methods do not fully exploit the complex structure of remotely sensed spectral images, ignoring the rich spatial information of the spectral images, which could boost the accuracy of these algorithms.

\subsection{SSC-based Methods for Land Cover Segmentation}
\label{sec:related_works_sub1}

Some SSC-based methods have been proposed for land cover segmentation, which take advantage of the neighboring spatial information but still present the scalability issue of SSC. Under the context of SIs, the $N_r\times N_c \times L$ 3D image data cube can be rearranged into a 2D matrix $\mathbf{X}\in \mathbb{R}^{D\times N}$ to apply the SSC algorithm, where $N=N_rN_c$ and $D<L$ is the number of features extracted from the spectral signatures after applying principal component analysis (PCA) \cite{sanchez2019supervised}. Taking into account that the spectral pixels belonging to the same land cover material are arranged in common regions, different works \cite{zhang2016spectral,zhai2016new,zhai2017kernel,bacca2017kernel,hinojosa2018spectral,hinojosa2018coded, hinojosa2021hyperspectral} aim at obtaining a piecewise smooth sparse coefficient matrix to incorporate such contextual dependence. In particular S-SSC \cite{zhang2016spectral} helps to guarantee spatial smoothness and reduce the representation bias by adding a regularization term in the SSC optimization problem which enforces a local averaging constraint on the sparse coefficient matrix. More recently, authors in \cite{hinojosa2021hyperspectral}, propose the 3DS-SSC algorithm which incorporate a 3D Gaussian filter in the optimization problem to perform a 3D convolution on the sparse coefficients, obtaining a piecewise-smooth representation matrix.

\section{Fast and Accurate Similarity-constrained Subspace Clustering (SC-SSC)}
\label{sec:prop_method}

\begin{figure*}[h]
	\begin{center}
		\includegraphics[width=\linewidth]{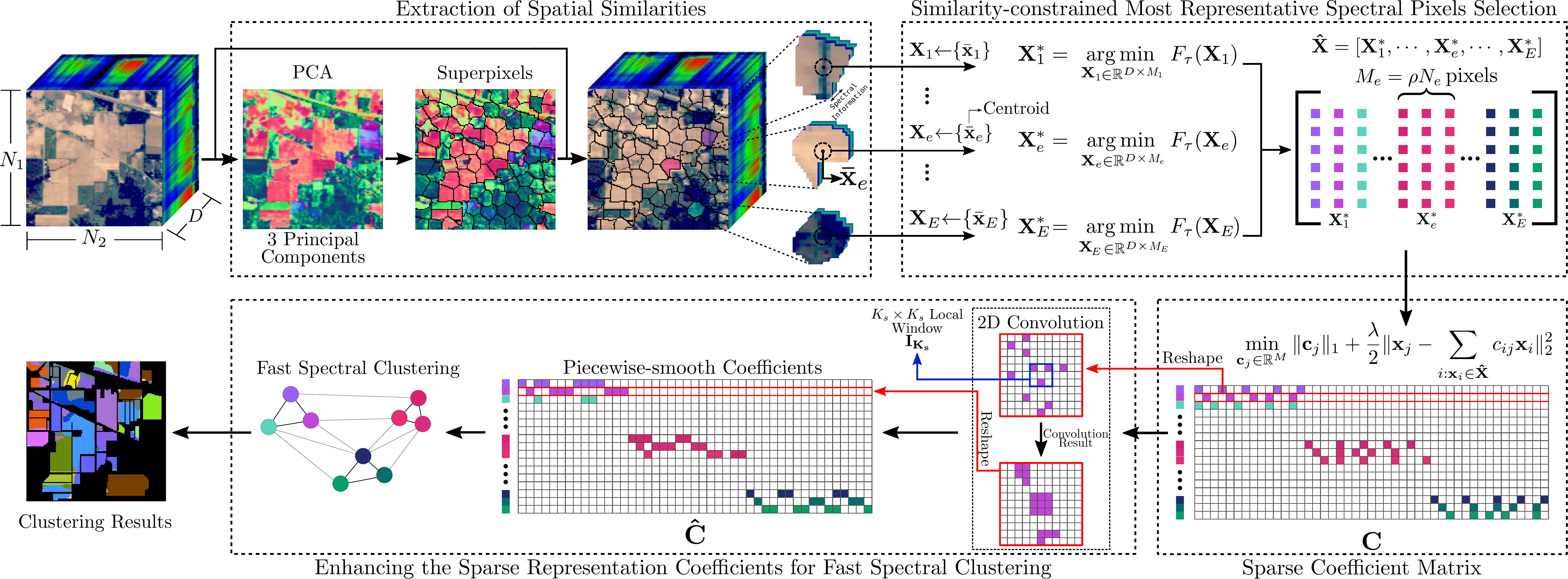}
	\end{center}
	\caption{Workflow of the proposed fast and accurate similarity-constrained subspace clustering algorithm (SC-SSC) for land cover segmentation. The overall algorithm is composed of four stages. In the first stage, we apply PCA to obtain the three principal components of the spectral image. Then, we segment the image in different subsets using a superpixel algorithm. Note that we only use PCA to extract spatial similarities, but we perform the following procedures on the spectral pixels, as we depicted with the circular flow symbols. In the second stage the most representative spectral pixels from each subset are obtained by solving Eq. (\ref{eq:findExmp_opt}) via Algorithm \ref{alg:selection}, and then all the representative spectral pixels from each subset are stacked as column in the matrix $\mathbf{\hat{X}}$. In the third stage, the $\left\{ \mathbf{c}_j \right\}$ vectors are obtained solving Eq. (\ref{eq:newSSC}). Finally, we reshape each row of the matrix $\mathbf{C}=\left[ \mathbf{c}_1,\cdots,\mathbf{c}_N \right]$, perform a 2D convolution with a $ K_s\times K_s $ kernel, and reshape back the result to obtain a piecewise-smooth coefficient matrix. We obtain the final data segmentation via fast spectral clustering, as described in Section \ref{subsec:enhanceSC}. The computational complexity of the overall algorithm is $O(\rho^2N^3)$, as analyzed on Section \ref{sec:comp_complexity}.}
	\label{fig:prop_method}
\end{figure*}

This section presents a subspace clustering algorithm for land cover segmentation that incorporates both properties: it can better handle large-scale datasets and takes advantage of the neighboring spatial information of SIs to boost the clustering accuracy. The complete workflow of the proposed method is shown in Fig. \ref{fig:prop_method}. In general, we exploit the self-representation property within subsets of neighboring similar pixels to select the most representative data points of the whole spectral image. Then, we enhance the sparse representation and perform fast spectral clustering to obtain the segmentation result.

\subsection{Similarity-constrained Most Representative Spectral Pixels Selection}

As neighboring spatial pixels commonly belong to the same land cover material, the proposed method aims to select a small subset of pixels that best represent their neighborhood. In this regard, we start by obtaining a segmentation map of the overall SI using a superpixels algorithm, which commonly expects a three-band image as input. Therefore, we first perform PCA to retrieve the three principal components of $\mathbf{X}$, and form the matrix $\mathbf{X}_{PCA}\in \mathbb{R}^{3\times N}$. Then, we use the SLIC algorithm \cite{achanta2012slic} to obtain a segmentation map $\mathbf{\tilde{m}} \in \mathbb{R}^{N}$ from $\mathbf{X}_{PCA}$, such that $\tilde{m}_j \in \left\{ 1,\cdots,E \right\}$, where $E$ is the number of segments. For instance, if $\tilde{m}_j=e$ means that the pixel $\mathbf{x}_j$ belongs to the segment $e$. Note that PCA is only performed to obtain $\mathbf{\tilde{m}}$ from $\mathbf{X}_{PCA}$ via SLIC; then, we use $\mathbf{\tilde{m}}$ to select the most representative spectral pixels $\mathbf{x}_j$ from $\mathbf{X}$ within each segment $e$.

Let $\mathbf{p}_e \in \mathbb{R}^{N_e}$ be the vector containing the indices of the $N_e$ most similar spectral pixels belonging to the subset $e$. We are interested in selecting the $M_e = \lfloor \rho N_e \rfloor$ most representative pixels from each subset, where $\rho \in (0,1)$. Taking advantage of the self-expressiveness property, the selection of the pixels within each neighborhood $e$ is obtained by searching for a subset $ \mathbf{X}_{e}^{*}  \subseteq \mathbf{X}$ that minimizes
\begin{equation}
	\mathbf{X}_{e}^{*} = \argmin_{ \mathbf{X}_e \in \mathbb{R}^{D\times M}} F_\tau (\mathbf{X}_e),
	\label{eq:findExmp_opt}
\end{equation}
where $F_\tau$ is the \textit{self-representation} cost function defined as
\begin{equation}
	F_{\tau}(\mathbf{X}_e) \coloneqq \sup_{\mathbf{x}_j \in \mathbf{X} \ :\ j \in \mathbf{p}_e} f_{\tau}(\mathbf{x}_j,\mathbf{X}_e). 
	\label{eq:self_rep_cost_f_subset}
\end{equation}
The metric function $f_{\tau}(\mathbf{x}_j,\mathbf{X}_e)$ geometrically measures how well a data point $\mathbf{x}_j \in \mathbf{X} : j \in \mathbf{p}_e$ can be represented by the subset $\mathbf{X}_e$, and we define it as
\begin{equation}
	f_{\tau}(\mathbf{x}_j,\mathbf{X}_e) \coloneqq \min_{\mathbf{c}_j \in \mathbb{R}^{N}} \|\mathbf{c}_j\|_1 + \frac{\tau}{2} \|\mathbf{x}_j - \sum_{i:\mathbf{x}_i \in \mathbf{X}_e} c_{ij}\mathbf{x}_i \|_2^2,
	\label{eq:ftau}
\end{equation}
where $\tau \in (1,\infty)$ is a parameter. Note that with Eq. (\ref{eq:self_rep_cost_f_subset}), we constrain Eq. \ref{eq:findExmp_opt} to search only for pixels $\mathbf{x}_j$ within the subset $e$, using the vector $\mathbf{p}_e$. To efficiently solve Eq. (\ref{eq:findExmp_opt}) for each subset $e$, we use the approximation algorithm described in Algorithm \ref{alg:selection}. Note that, instead of using a random initialization, we select the centroid spectral pixel $\mathbf{\bar{x}}_e$ as the initialization data point since it is the most similar point, in the Euclidean distance, to all other data points in $e$. The search space constraint--given by dividing the SI into subsets--in conjunction with selecting the centroid spectral pixel speeds up the acquisition of the most representative spectral pixels.

\begin{algorithm}[t]
	\SetKwFunction{isOddNumber}{isOddNumber}
	\SetKwInOut{KwIn}{Input}
	\SetKwInOut{KwOut}{Output}
	\KwIn{Data $\mathbf{X} \in \mathbb{R}^{D\times N}$, Indices vector $\mathbf{p}_e \in \mathbb{R}^{N_e}$, Parameters $0<\rho <1$, and $\tau >1$.}
	\KwOut{$\mathbf{X}_e \in \mathbb{R}^{D \times \lfloor \rho N_e \rfloor}$.}
	
	\SetAlgoLined
	\SetKwProg{Fn}{Function}{}{end}
	\nonl \Fn{Data\_Selection($\mathbf{X},\mathbf{p}_e, \rho, \tau$)}{
		
		$\mathbf{\bar{x}}_e \leftarrow \text{centroid}(\{\mathbf{x}_j \in \mathbf{X}:j \in \mathbf{p}_e\})$
		
		$\mathbf{X}_e^{(1)} \leftarrow \left\{ \mathbf{\bar{x}}_e \right\}$
		
		\mycommfont{$\triangleright$ $(\mathbf{p}_e)_{k}$ gets the $k$ element of the vector $\mathbf{p}_e$.}
		
		Compute $b_k = f_\tau (\mathbf{x}_j,\mathbf{X}_{e}^{(1)})$ for $k=1, \cdots,N_e$, and $j=(\mathbf{p}_e)_{k}$.
		
		$M_e \leftarrow \lfloor \rho N_e \rfloor$
		
		\For{$i=1,\cdots,M_e-1$}{
			Let $o_1,\cdots,o_{N_e}$ be an ordering of $1,\cdots,N_e$ such that $b_{o_p} \ge b_{o_q}$ when $p<q$.
			
			Initialize $\textit{max\_cost}=0$.
			
			\For{$k=1,\cdots,N_e$}{
				
				Set $b_{o_k}=f_\tau(\mathbf{x}_{o_k},\mathbf{X}_{e}^{(i)}).$
				
				\If{$b_{o_k} > \textit{max\_cost}$}{
					Set $\textit{max\_cost}=b_{o_k}$, and $\textit{new\_index}=o_k$.
				}
				
				\If{$k=N_e$ or $\textit{max\_cost} \ge b_{o_{k+1}}$}{
					\textbf{break}
				}
			}
			$\mathbf{X}_{e}^{(i+1)} = \mathbf{X}_{e}^{(i)} \cup \left\{ \mathbf{x}_{\textit{new\_index}} \right\}$
		}
		\KwRet{$\mathbf{X}_{e}$}
	}
	\caption{Similarity-constrained spectral pixels selection}
	\label{alg:selection}
\end{algorithm}

\subsection{Enhancing the sparse representation coefficients for fast spectral clustering}
\label{subsec:enhanceSC}

Once the most representative spectral pixels from each subset are obtained, we build the matrix $\mathbf{\hat{X}}$ by stacking the results as columns, i.e., $\mathbf{\hat{X}}=\left[ \mathbf{X}_1,\cdots,\mathbf{X}_E \right]$. Then, the sparse coefficient matrix $\mathbf{C}$ of size $M \times N$, with $M = \lfloor \rho N \rfloor$, can be obtained by solving the following optimization problem, similar to Eq. \ref{eq:ftau},
\begin{equation}
	\min_{\mathbf{c}_j \in \mathbb{R}^M} \|\mathbf{c}_j\|_1 + \frac{\tau}{2}\|\mathbf{x}_j - \sum_{i:\mathbf{x}_i \in \mathbf{\hat{X}}} c_{ij}\mathbf{x}_i \|_2^2, \quad \forall \ \mathbf{x}_j \in \mathbf{X}. 
	\label{eq:newSSC}
\end{equation}
Note that $\mathbf{C}$ encodes information about the similarities between $\mathbf{\hat{X}}$ and $\mathbf{X}$. Besides, each row of $\mathbf{C}$ contains the representation coefficients distribution of the whole image with respect to a single representative pixel. Taking into account that spectral pixels belonging to the same land cover material should be regionally distributed in the image, i.e., two spatially neighboring pixels in a SI usually have a high probability of belonging to the same class. Then, according to the self-expressiveness property, their representation coefficients should also be very close concerning the same sparse basis; hence, each row of $\mathbf{C}$ should be piecewise-smooth. Therefore, an intuitive approach to include the spatial information to boost the clustering performance is to apply a 2D smoothing convolution on the sparse coefficients $\mathbf{C}$. Given a blur kernel matrix $\mathbf{I}_{K_s}$ of size $K_s \times K_s$, we will denote the 2D convolution process as $\mathbf{\hat{C}} = \mathcal{G}(\mathbf{C},\mathbf{I}_{K_s})$. Specifically, as depicted in Fig. \ref{fig:prop_method} within dashed blue line, we propose to perform $\mathcal{G}$ by first reshaping each row of $\mathbf{C}$ to a window of size $N_r\times N_c$, which corresponds to the spatial dimensions of the SI, and then conducting the convolution with $\mathbf{I}_{K_s}$. Finally, the convolution result is rearranged back as a row vector of the piecewise-smooth coefficient matrix $\mathbf{\hat{C}}=\left[ \mathbf{\hat{c}}_1,\cdots,\mathbf{\hat{c}}_N \right] \in \mathbb{R}^{M\times N}$.

\begin{algorithm}[t]
	\SetKwFunction{isOddNumber}{isOddNumber}
	
	\SetKwComment{tcp}{$\triangleright$ }{}%
	\SetKwInOut{KwIn}{Input}
	\SetKwInOut{KwOut}{Output}
	
	\KwIn{The spectral image in matrix form $\mathbf{X} \in \mathbb{R}^{D\times N}$, parameters $\tau >1$, $0<\rho <1$, $E > 1$, $K_s>1$.}
	\KwOut{The segmentation of $\mathbf{X}$.}
	
	$\mathbf{X}_{PCA}\in \mathbb{R}^{3\times N}\leftarrow PCA(\mathbf{X})$
	
	$\mathbf{\tilde{M}} \leftarrow Superpixels(\mathbf{X}_{PCA})$
	
	$\mathbf{\tilde{m}} \leftarrow \text{vec}(\mathbf{\tilde{M}})$
	
	$\mathbf{\hat{X}}^{(e)} \leftarrow \emptyset$
	
	\For{$e \leftarrow 1$ \KwTo $E-1$}{
		$\mathbf{p}_e = \left\{ j : \mathbf{\tilde{m}}_j = e, \forall j\in \{1,\cdots,N\} \right\} $
		
		$\mathbf{X}_{e} \leftarrow Data\_Selection(\mathbf{X},\mathbf{p}_e, \rho, \tau)$
		
		$\mathbf{\hat{X}}^{(e+1)} = \mathbf{\hat{X}}^{(e)} \cup \mathbf{X}_e$
	}
	
	Compute $\mathbf{C} = \left[ \mathbf{c}_1,\cdots,\mathbf{c}_N \right]$ by solving Eq. (\ref{eq:newSSC}).

	$\mathbf{I}_{K_s} = (1/K_s^2)\cdot\mathbf{1} \quad$ \tcp{where $\mathbf{1}$ is an  all-ones matrix of size $K_s$}
	
	
	$\mathbf{\hat{C}} = \mathcal{G}(\mathbf{C},\mathbf{I}_{K_s})$ 
	
	$\mathbf{\tilde{C}}=\left[ \mathbf{\hat{c}}_1 / \|\mathbf{\hat{c}}_1\|_2, \cdots, \mathbf{\hat{c}}_N / \|\mathbf{\hat{c}}_N\|_2 \right]$
	
	$\mathbf{\alpha} = \sum_{j=1}^{N} \mathbf{\tilde{c}}_j$
	
	$\mathbf{D} = \text{diag}(\mathbf{\tilde{C}}^T \mathbf{\alpha})$
	
	Run $k$-means clustering algorithm on the top $k$ right singular vectors of $\mathbf{\tilde{C}D}^{-1/2}$ to obtain the segmentation of $\mathbf{X}$.
	
	\KwRet{The cluster assignments of $\mathbf{X}$}
	\caption{SC-SSC for land cover segmentation}
	\label{alg:SC-SSC_all}
\end{algorithm}

Since $\mathbf{\hat{C}}$ is not square, it is not feasible to directly build the affinity matrix $\mathbf{A}$ be used with spectral clustering as in SSC \cite{von2007tutorial,elhamifar2013sparse}. To resolve this issue we use a fast spectral clustering approach to efficiently obtain the spectral embedding of the input data. Specifically, let us consider the columns of $\mathbf{\tilde{C}}=\left[ \mathbf{\tilde{c}}_1,\cdots,\mathbf{\tilde{c}}_N \right] \in \mathbb{R}^{M \times N}$, where $\mathbf{\tilde{c}}_j=|\mathbf{\hat{c}}_j|/\|\mathbf{\hat{c}}_j\|_2$, and compute the $i$-th element of the degree matrix $\mathbf{D}$ as follows
\begin{equation}
	(\mathbf{D})_i = \sum_{j=1}^{N} A_{ij} = \sum_{j=1}^{N} \mathbf{\tilde{c}}_i^T \mathbf{\tilde{c}}_j = \mathbf{\tilde{c}}_i^T \sum_{j=1}^{N} \mathbf{\tilde{c}}_j = \text{diag}(\mathbf{\tilde{C}}^T\boldsymbol{\alpha})_{i},
\end{equation}
where $\boldsymbol{\alpha} = \sum_{j=1}^{N} \mathbf{\tilde{c}}_j \in \mathbb{R}^{M}$. Next, we can find the eigenvalue decomposition of $\mathbf{D}^{-1/2}\mathbf{AD}^{-1/2}$ by computing the singular value decomposition \cite{golub1971singular} of $\mathbf{\tilde{C}D}^{-1/2} \in \mathbb{R}^{M \times N}$. Finally, the segmentation of the data can be obtained by running the $k$-means algorithm on the top $k$ right singular vectors for $\mathbf{\tilde{C}D}^{-1/2}=\mathbf{U \Sigma P}^T$. As a result, the computational complexity of spectral clustering in our framework is linear with respect to the size of the data $N$, which makes it suitable for large-scale datasets. The proposed SC-SSC method is summarized in Algorithm \ref{alg:SC-SSC_all}, and its computational complexity is analyzed on Section \ref{sec:comp_complexity}.

\subsection{Analysis of the Proposed Method}

We now analyze how the proposed method optimizes the sparsity (subspace-preserving property) and the connectivity in the representation coefficient matrix. Furthermore, we analyze the computational complexity of Algorithm \ref{alg:SC-SSC_all}.

\subsubsection{Subspace-preserving Property and Connectivity}

As mentioned in Section \ref{sec:related_works}, one of the main requirements for the success of subspace clustering methods is that the optimization process recovers a subspace-preserving solution. Specifically, the non-zero entries of the sparse representation vector $\mathbf{c}_j$ should be related only to the intra-subspace samples of $\mathbf{x}_j$. Indeed, as the following definition states, the representation coefficients among intra-subspace data points are always larger than those among inter-cluster points.

\theoremstyle{definition}
\begin{definition}[\small \textbf{Intra-subspace projection dominance, IPD \cite{peng2016constructing}}]
	The IPD property of a coefficient matrix $\mathbf{C}$ indicates that for all $\mathbf{x}_u,\mathbf{x}_v \in \mathcal{S}$ and $\mathbf{x}_q \notin \mathcal{S}$, where $u,v,q \in \left\{1,\cdots,N\right\}$, and $\mathcal{S}$ is a subspace of $\mathbf{X}$, we have $C_{uv} \ge C_{uq}$.
	\label{def:IPD}
\end{definition}


Since the proposed method selects the most representative spectral pixels for each subset $e$ based on the self-representation property, it is expected that each subset is subspace-preserving, i.e., $c_{ij}$ is nonzero only if $\mathbf{x}_i$ and $\mathbf{x}_j$, for $i,j \in \mathbf{p}_e$, belong to the same subspace $\mathcal{S}$. Furthermore, note that it is very probable that a subset $e$ has more spectral pixels from the same class due to the spatial dependence in SI; then, the resulting coefficients vector will have large values for those spectral pixels within $e$. Therefore, the strategy adopted in the proposed method will improve the structure of the vectors $\mathbf{c}_j$ obtained by Eq. (\ref{eq:newSSC}) and will improve the probability that $\mathbf{c}_j$ satisfies the IPD.

Besides, using the 2D smoothing convolution procedure $\mathcal{G}(\mathbf{C},\mathbf{I}_{K_s})$, the proposed method improves the connectivity of the data points by preserving the most significant values in the coefficient matrix $\mathbf{C}$ and reducing the small or noisy isolated values, based on the IPD property \cite{peng2016constructing}. Then, the resulting matrix $\mathbf{\hat{C}}$ will have localized neighborhoods in the sparse codes making the representation coefficients of spatially neighboring pixels very close as well, following our main assumption in section \ref{subsec:enhanceSC}.

\subsubsection{Computational Complexity Analysis}
\label{sec:comp_complexity}
As shown in Fig. \ref{fig:prop_method}, the proposed method mainly involves four stages: the extraction of spatial similarities, the selection of similarity-constrained representative spectral pixels, the sparse coefficient matrix estimation by solving Eq. (\ref{eq:newSSC}), and enhancing the representation coefficients for fast spectral clustering. Given a spectral image in matrix form $\mathbf{X}\in \mathbb{R}^{D\times N}$ and $E$ subsets $\mathbf{X}_e \subseteq \mathbf{X}$ of dimensions $D \times M_e$, with $M_e = \rho N_e$, we will show the complexity of each stage before establishing the total complexity of Algorithm \ref{alg:SC-SSC_all}. Specifically, in the first stage, we acquire the segmentation map $\mathbf{\tilde{m}}$ for an SI. Such procedure involves computing PCA over $\mathbf{X}$ to retrieve only the three principal components, which takes $O(N)$, and performing SLIC superpixels \cite{achanta2012slic} which also has linear time complexity $O(N)$. The second stage requires to execute Algorithm \ref{alg:selection}, which has $O(\rho N_e^2)$ time complexity over $E$ subsets, then the overall complexity of this stage will be $O(\rho \max(N_1^2,\cdots,N_E^2))$. The third stage entails solving Eq. (\ref{eq:newSSC}) which is a LASSO problem that can be efficiently computed in $O(M^2N)$ using the LARS algorithm \cite{efron2004least}. Finally, in the last stage, the 2D convolution takes $O(N)$ as $K_s \ll N$ and, since for the spectral clustering we only need the $k$ largest singular values, we can use the truncated singular value decomposition (SVD), which takes $O(k^2N)$. Thus, the overall complexity of this stage is $O(k^2N)$. Therefore, the complexity of Algorithm \ref{alg:SC-SSC_all} will be dominated by the complexity of the third stage, hence it will run in $O(M^2N)=O(\rho^2N^3)$, where $\rho \in (0,1)$.

\begin{figure}[h]
	\centering
	\includegraphics[width=\columnwidth]{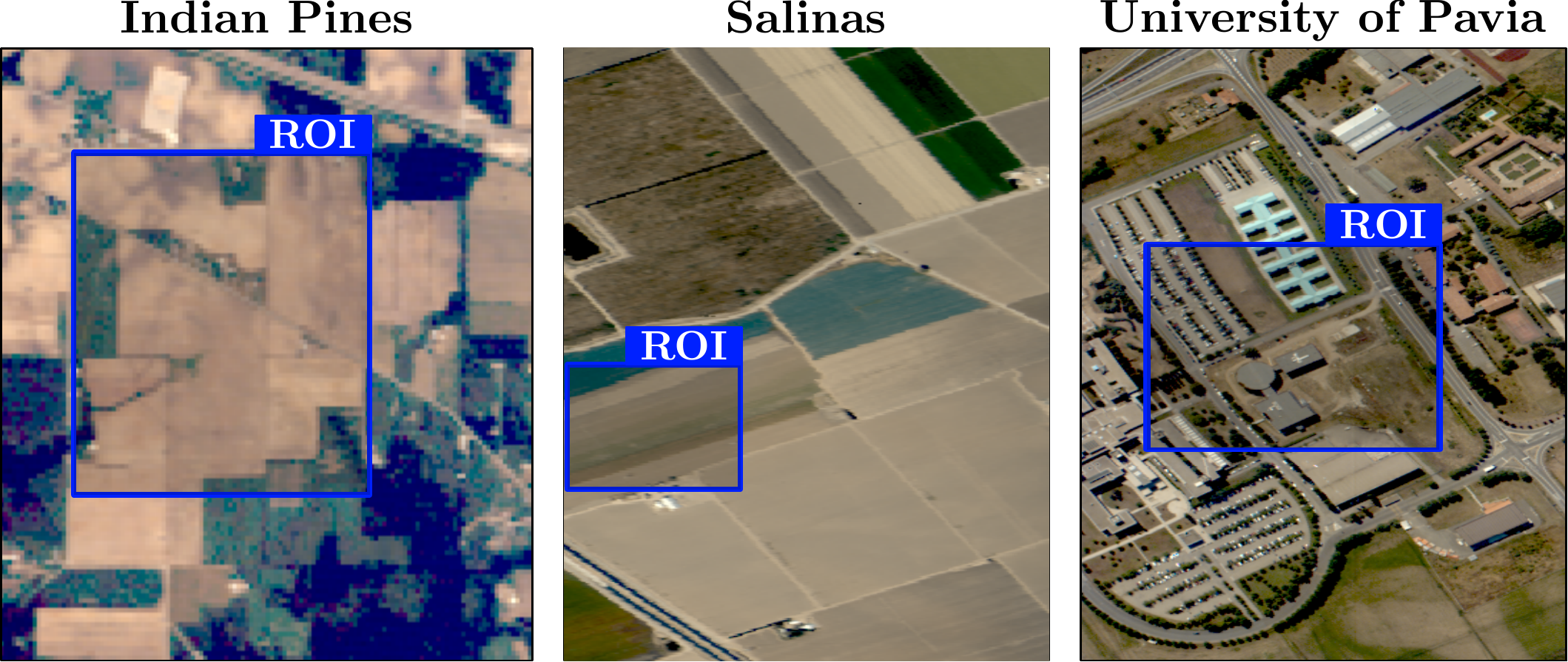}
	\caption{False-color images and regions of interest (ROI) for the three real remote sensing images used in the experiments.}
	\label{fig:datasets}
\end{figure}

\section{Experimental Evaluation}
\label{sec:experiments}

In this section we show the performance of SC-SSC\footnote{A MatLab implementation of the Algorithm \ref{alg:SC-SSC_all} can be found at \url{https://link.carloshinojosa.me/SC-SSC}.} for land cover segmentation. The sparse optimization problem in Eq. (\ref{eq:ftau}), and Eq. (\ref{eq:newSSC}) are solved by the LASSO version of the LARS algorithm \cite{efron2004least} implemented in the SPAMS package \cite{mairal2010online}. All the experiments were run on an Intel Core i7 9750H CPU (2.60GHz, 6 cores), with $32$ GB of RAM.

\subsection{Setup}

\textbf{Databases}. The proposed subspace clustering approach (SC-SSC) was tested on three well-known hyperspectral images\footnote{\url{http://www.ehu.eus/ccwintco/index.php?title=Hyperspectral_Remote_Sensing_Scenes}.} with different imaging environments, see Fig. \ref{fig:datasets}. The \textbf{Indian Pines} hyperspectral data set has $145\times 145$ pixels and $200$ spectral bands in the range of $0.4-2.5 \mu m$. The second scene, \textbf{Salinas}, has $512 \times 217$ pixels and $204$ spectral bands in the range of $0.24-2.40 \mu m$. The third scene, \textbf{University of Pavia}, comprises $610 \times 340$ pixels, and has $103$ spectral bands with spectral coverage ranging from $0.43-0.84 \mu m$. In order to make a fair comparison with non-scalable methods, we select, for each image, the most frequently used region of interest (ROI) in spectral image clustering, as shown in Fig. \ref{fig:datasets}. The Indian Pines ROI has a size of $70\times 70$ pixels, which includes four main land-cover classes: corn-no-till, grass, soybeans-no-till, and soybeans-min-till. The Salinas ROI comprises $83 \times 83$ pixels and includes six classes: brocoli-1, corn-senesced, lettuce-4wk, lettuce-5wk, lettuce-6wk, and lettuce-7wk. Finally, the University of Pavia ROI is composed of $200\times 200$ pixels, and includes all the classes (nine) as in the full image: asphalt, meadows, gravel, trees, metal sheets, bare soil, bitumen, bricks, and shadows. For all experiments (including the baseline methods), we reduce the spectral dimensions of each image using PCA to $D=0.25L$, where $L$ is the number of spectral bands. Then, we rearrange the data cube to form a matrix $\mathbf{X} \in \mathbb{R}^{D\times N}$, and normalize the columns (spectral pixels) to have unit $\ell_2$ norm.

\textbf{Baselines and Evaluation Metrics.} We compare our approach with the SSC-based methods highlighted in Section \ref{sec:related_works_sub2}: SSC\cite{elhamifar2013sparse}, SSSC \cite{peng2013scalable}, SSC-OMP \cite{you2016scalable}, ORGEN \cite{you2016oracle}, SR-SSC \cite{abdolali2019scalable}, ESC-FFS \cite{you2018scalable}, S-SSC \cite{zhang2016spectral}, and 3DS-SSC \cite{hinojosa2021hyperspectral}. We also show the results with SSC  as an additional reference. To make a fair comparison, we compare the performance of the proposed SC-SSC algorithm with the non-scalable methods (SSC, S-SSC, 3DS-SSC, and ORGEN) on the ROIs of the remote sensing images shown in Fig. \ref{fig:datasets}. Then, we compare the performance of the SC-SSC with the scalable methods (SSC-OMP, ESC-FFS, SR-SSC, and SSSC) on the full hyperspectral images. For the sake of completeness, we also compare our approach with non-SSC-based methods that use fast spectral clustering \cite{wei2019fast,wang2017fast,wang2019scalable}. Specifically, we gather the clustering results on the Indian Pines image reported on such works and compared them against the proposed method. To compare the clustering performance of our model, we rely on five standard metrics: user's accuracy (UA), average accuracy (AA), overall accuracy (OA), Kappa coefficient, and normalized mutual information (NMI) \cite{lillesand2015remote,strehl2002cluster}. In particular, UA, AA, OA, and Kappa coefficient can be obtained by means of an error matrix (a.k.a confusion matrix) \cite{lillesand2015remote}. UA represents the clustering accuracy of each class, while AA is the mean of UA, and OA is computed by dividing the total number of correctly classified pixels by the total number of reference pixels. UA, AA, and OA values are presented in percentage, while Kappa coefficients and NMI values range from 0 (poor clustering) to 1 (perfect clustering). We also compare the methods in terms of clustering time, and the results are presented in seconds.


\subsection{Parameters analysis and tuning}

\begin{figure}[t]
	\centering
	\hspace{-0.2in}\includegraphics[width=0.75\columnwidth]{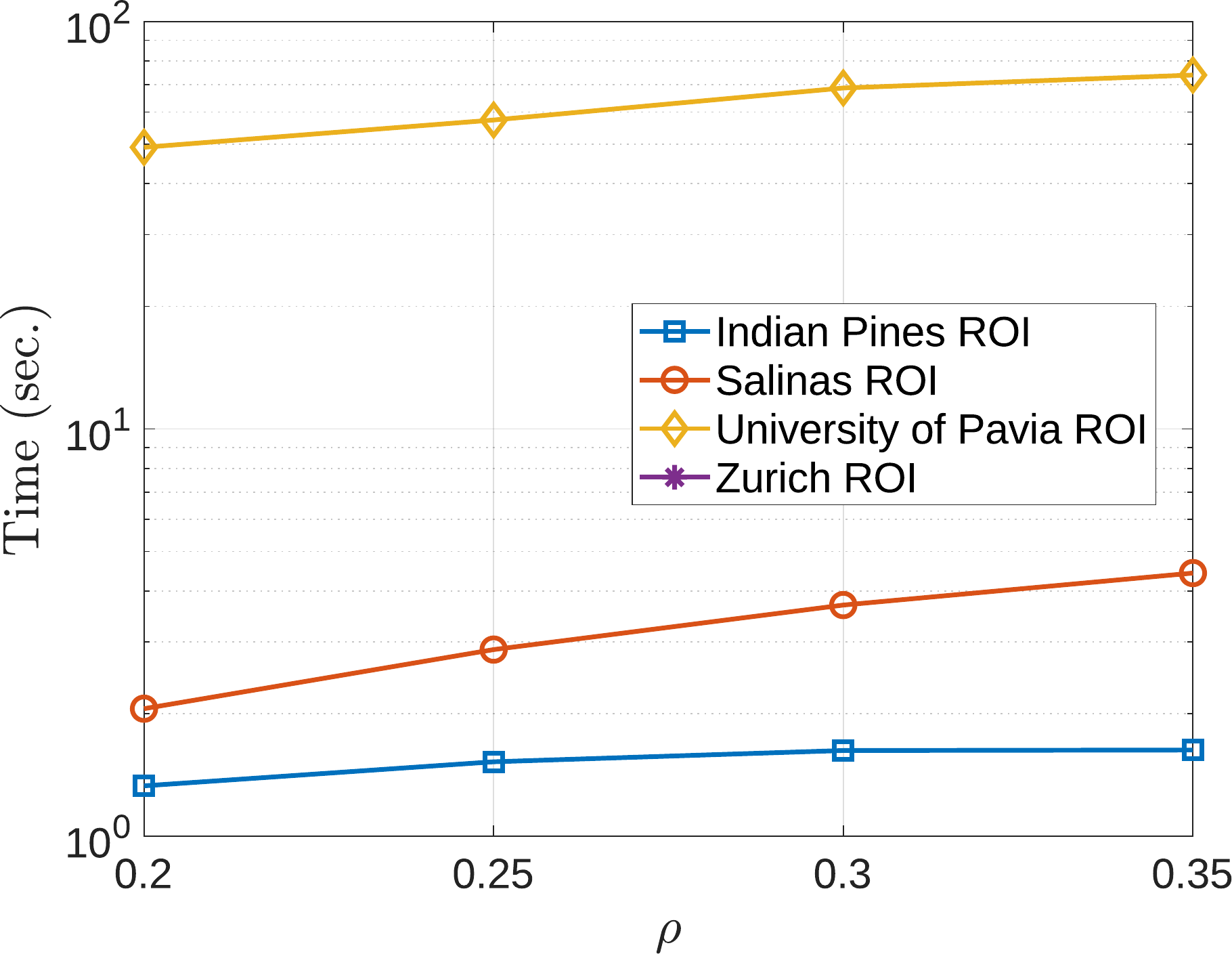}
	\caption{Running time (seconds) as a function of $\rho$.}
	\label{fig:rho_plot}
\end{figure}

The parameters $\rho, E$, and $K_s$ of Algorithm \ref{alg:SC-SSC_all} were manually adjusted for each dataset. We conduct different experiments varying each parameter, with the others fixed, to obtain the best overall accuracy with each spectral image. During simulations, we observe that the parameter $\rho$ has a direct impact on the execution time of the proposed method. Figure \ref{fig:rho_plot} presents the running time of SC-SSC for all the databases. As shown, increasing $\rho$ directly increases the running time; however, the most significant increment in time is given by the number of spectral pixels $N$, as observed with the differences in time between the curves. As we analyze in Section \ref{sec:comp_complexity}, this behavior is expected since the computational complexity of the algorithm is $O(\rho^2N^3)$. 

\begin{table}[t]
	\centering
	{\footnotesize \caption{Selected parameters in Algorithm \ref{alg:SC-SSC_all} for each testing spectral images.}
		\label{tab:parameters}}
	\resizebox{0.82\columnwidth}{!}{%
		\begin{tabular}{cccc}
			\hline
			\textbf{Parameter} & \textbf{Indian Pines} & \textbf{Salinas} & \textbf{University of Pavia} \\ \hline
			$\rho$             & 0.35                  & 0.2              & 0.3                          \\
			$E$                & 1700                  & 700              & 1900                         \\
			$K_s$              & 8                     & 3                & 8                            \\ \hline
		\end{tabular}
	}
\end{table}

\begin{figure*}
	\centering
	\includegraphics[width=\linewidth]{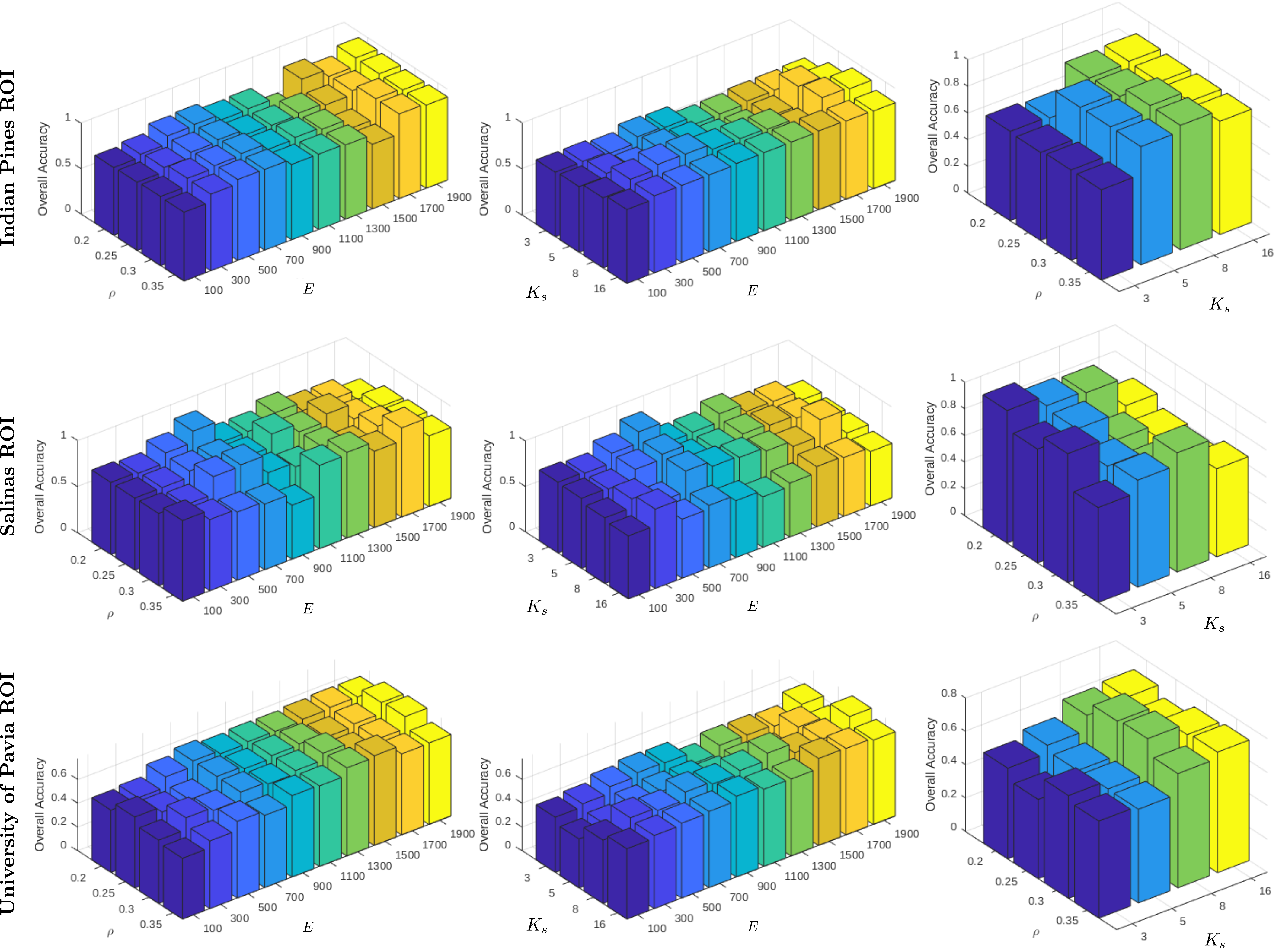}
	\caption{Experimental results of different combinations of parameters $\rho,E$, and $K_s$ for a fixed $\lambda$. Each row presents the 3D bar plot of $\rho$ vs. $E$, $K_s$ vs. $E$, and $\rho$ vs. $K_s$ for each database,  and the evaluation is given by the overall accuracy with values between 0 and 1. The plot $\rho$ vs. $E$ shows how the OA changes when the number of selected representative data points varies concerning the amount of superpixels $E$. $K_s$ vs. $E$ depicts how the OA is affected by the amount of superpixels and the kernel size used in the 2D convolution to enhance the sparse coefficient matrix. Finally, $\rho$ vs. $K_s$ shows the change in OA when the number of selected representative data points varies, and a specific kernel size is used in the 2D convolution. The three plots prove the importance of an adequate balance between the selection of the number of representative pixels and the inclusion of spatial information in the spectral clustering algorithm.}
	\label{fig:all_param_propM}
\end{figure*}

To find the best configuration, the parameters were varied between the following values: $\tau \in \left\{5,10,15,20\right\}$, $\rho \in \left\{ 0.2,0.25,0.3,0.35 \right\}$, $E \in \{ 100,300,500,700,900,$ $1100,1300,1500,1700,1900 \}$, $K_s \in \left\{ 3,5,8,16 \right\}$. Figure \ref{fig:all_param_propM} shows the performance of the proposed method with a different combination of the parameters for all the databases, where the overall accuracy is shown between 0 and 1, and the parameter $\tau$ was fixed. Given the results of the different combinations of the parameters, we selected the best ones and summarized them in Table \ref{tab:parameters}. By analyzing Figure \ref{fig:all_param_propM}, we observe that the precision changes with different values of $\rho$ and $K_s$. The parameter $\rho$ determines the number of the selected most-representative data points within each of the $E$ segments, and $K_s$ is the kernel size used in the 2D convolution. Then, we can conclude that an adequate balance between the selection of the number of representative pixels and the inclusion of spatial information in the spectral clustering algorithm is crucial to obtain the best performance.

In order to make a fair comparison, we also performed several simulations with the baseline methods to manually select the best parameters in their configurations. Table \ref{table:selectParamsOM} presents the selected parameters for each method after running the experiments. We present the same parameters symbols used in the original works. Note that, for the SSC, S-SSC, and 3DS-SSC algorithms, we select the same parameters reported in \cite{hinojosa2018coded} and \cite{hinojosa2021hyperspectral}, since they are already optimal for spectral image clustering. 



\begin{table}[t]
	\centering
	{\footnotesize \caption{Selected parameters for the baseline methods.}
	\label{table:selectParamsOM}}
	\resizebox{0.92\columnwidth}{!}{%
		\begin{tabular}{@{}lccc@{}}
			\toprule
			\textbf{}        & \textbf{Indian Pines}               & \textbf{Salinas}                     & \textbf{University of Pavia}         \\ \midrule
			\textbf{ESC-FFS \cite{you2018scalable}} & $\lambda$=10, k =700, t = 10.       & $\lambda$=20, k=700, t=10.           & $\lambda$=15, k=700, t=20.           \\ \midrule
			\textbf{SR-SSC \cite{abdolali2019scalable}}  & \begin{tabular}[c]{@{}c@{}}$\alpha$=200, nGraph=5,\\ Nsample=10.\end{tabular}
			& \begin{tabular}[c]{@{}c@{}}$\alpha$=300, nGraph=20,\\ Nsample=20.\end{tabular} & \begin{tabular}[c]{@{}c@{}} $\alpha$=140, nGraph=15,\\ Nsample=10.\end{tabular} \\ \midrule
			\textbf{ORGEN \cite{you2016oracle}}   &\begin{tabular}[c]{@{}c@{}} $\lambda$=0.8, nu=10,\\ Nsample=200.\end{tabular} &\begin{tabular}[c]{@{}c@{}} $\lambda$=0.95, nu=50,\\ Nsample=400.\end{tabular} & \begin{tabular}[c]{@{}c@{}}$\lambda$=0.8, nu=100,\\ Nsample=400.\end{tabular} \\ \midrule
			\textbf{SSSC \cite{peng2013scalable}}    & $\lambda$=0.01, tol=0.0001.         & $\lambda$=0.001, tol=0.001.          & $\lambda$=1e-06, tol=0.01.           \\ \midrule
			\textbf{SSC-OMP \cite{you2016scalable}} & K=40, thr=1e-07                     & K=30, thr=0.001                     & K=10, thr=1e-06                      \\ \bottomrule
		\end{tabular}
	}
\end{table}

\begin{table}[t]
	{\footnotesize \caption{Ablation study of our method. The configuration shown in bold (Experiment IV) corresponds to our proposed approach which leads to the best results in terms of OA and NMI.}
	\label{table:ablations}}
	\resizebox{\columnwidth}{!}{%
		\setlength{\tabcolsep}{3pt}
		\begin{tabular}{cccccccccc}
			&            &             &                                 & \multicolumn{2}{c}{Indian Pines}                   & \multicolumn{2}{c}{Pavia}                           & \multicolumn{2}{c}{Salinas}    \\ \hline
			\multicolumn{1}{c|}{Experiment}  & PCA        & Superpixels & \multicolumn{1}{c|}{2D Conv}    & OA             & \multicolumn{1}{c|}{NMI}          & OA             & \multicolumn{1}{c|}{NMI}           & OA             & NMI           \\ \hline
			\multicolumn{1}{c|}{I}           & \checkmark          & \checkmark           & \multicolumn{1}{c|}{}           & 64.36          & \multicolumn{1}{c|}{0.36}         & 37.78          & \multicolumn{1}{c|}{0.49}          & 74.26          & 0.65          \\
			\multicolumn{1}{c|}{II}          &            & \checkmark           & \multicolumn{1}{c|}{\checkmark}          & 80.53          & \multicolumn{1}{c|}{0.63}         & 67.04          & \multicolumn{1}{c|}{0.81}          & 84.63          & 0.86          \\
			\multicolumn{1}{c|}{III}         &            &             & \multicolumn{1}{c|}{\checkmark}          & 41.72          & \multicolumn{1}{c|}{0.29}         & 38.11          & \multicolumn{1}{c|}{0.46}          & 43.33          & 0.42          \\
			\multicolumn{1}{c|}{IV}          & \checkmark          &             & \multicolumn{1}{c|}{\checkmark}          & 41.81          & \multicolumn{1}{c|}{0.3}          & 38.21          & \multicolumn{1}{c|}{0.47}          & 49.53          & 0.41          \\
			\multicolumn{1}{c|}{V}           &            & \checkmark           & \multicolumn{1}{c|}{}           & 65.84          & \multicolumn{1}{c|}{0.37}         & 47.60          & \multicolumn{1}{c|}{0.49}          & 73.29          & 0.77          \\
			\multicolumn{1}{c|}{\textbf{VI}} & \textbf{\checkmark} & \textbf{\checkmark}  & \multicolumn{1}{c|}{\textbf{\checkmark}} & \textbf{93.14} & \multicolumn{1}{c|}{\textbf{0.8}} & \textbf{77.57} & \multicolumn{1}{c|}{\textbf{0.82}} & \textbf{99.42} & \textbf{0.98} \\ \hline
	\end{tabular}}
\end{table}

\subsection{Ablation Studies}
We conduct six ablation experiments to investigate different configurations for the proposed subspace clustering approach. Specifically, we compare the proposed workflow's performance in Fig. \ref{fig:prop_method} when incorporating/excluding PCA, superpixels, and the 2D convolution. Table \ref{table:ablations} present the results obtained from the different combinations in terms of OA and NMI for the three tested images. We observed that using superpixels to extract spatial similarities improves the clustering performance for the three tested images in all the cases, which evidence the importance of the neighboring spatial information in our workflow. Also, using superpixels and the 2D convolution (Experiment II) leads to the second-best result, while only using 2D convolution (Experiment III) does not lead to a significant clustering improvement. Finally, Experiment VI corresponds to our proposed approach where we show that we achieve the best results in terms of OA and NMI when using the three operations as described in the workflow in Fig. \ref{fig:prop_method}.

\subsection{Visual and Quantitative Results}
\label{sec:experiments_VQ}

\subsubsection{Comparison with non-scalable methods} 
\begin{figure*}
	\centering
	\includegraphics[width=\linewidth]{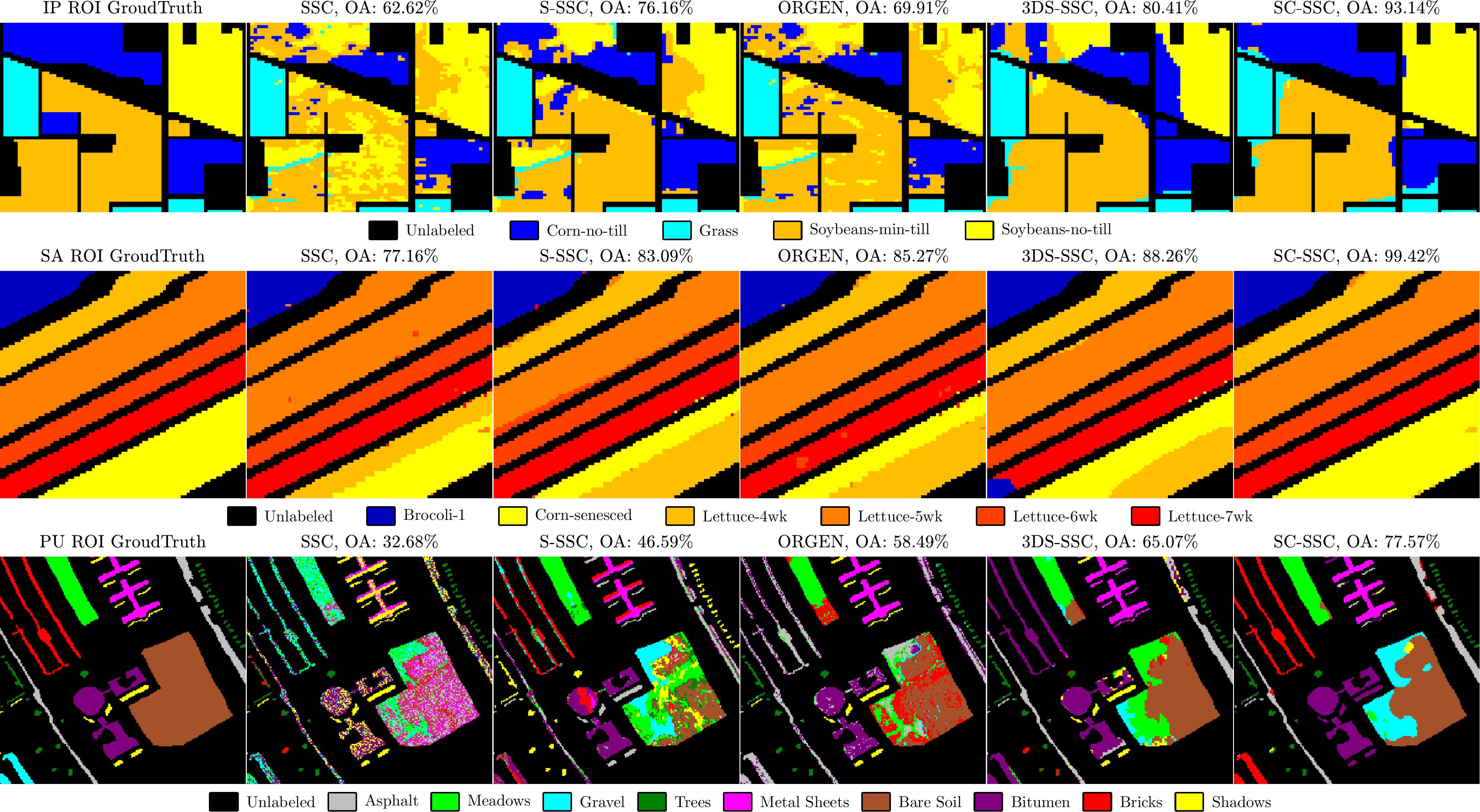}
	\caption{Land cover maps of (first row) Indian Pines ROI, (second row) Salinas ROI, and (last row) University of Pavia ROI. The proposed method is compared with the methods that perform best on these spectral images.}
	\label{fig:visual_results}
\end{figure*}

Figure \ref{fig:visual_results} presents the obtained land cover maps on the Indian Pines, Salinas, and University of Pavia ROIs, where we compare the performance of our SC-SSC method with the non-scalable methods: SSC, S-SSC, ORGEN, and 3DS-SSC. The quantitative evaluations corresponding to the UA, AA, OA, Kappa, NMI, and Time with the non-scalable clustering methods are reported in Table \ref{table:QuantRes}, in which the best results are shown in bold and the second-best is underlined. From Table \ref{table:QuantRes}, it can be clearly observed that, in general, the proposed SC-SSC method performs better than others. Specifically, SC-SSC achieves an OA of $93.14\%$ and $99.42\%$, in only $1.63$ and $2.06$ seconds, for the Indian Pines and Salinas dataset, respectively, which are remarkable results for unsupervised learning settings. Similarly, for the University of Pavia ROIs, it is observed from Table \ref{table:QuantRes} that the proposed SC-SSC achieves the best clustering performance in all the accuracy evaluation metrics, among all the other algorithms.

\subsubsection{Comparison with scalable methods} 
\begin{figure*}
	\centering
	\includegraphics[width=\linewidth]{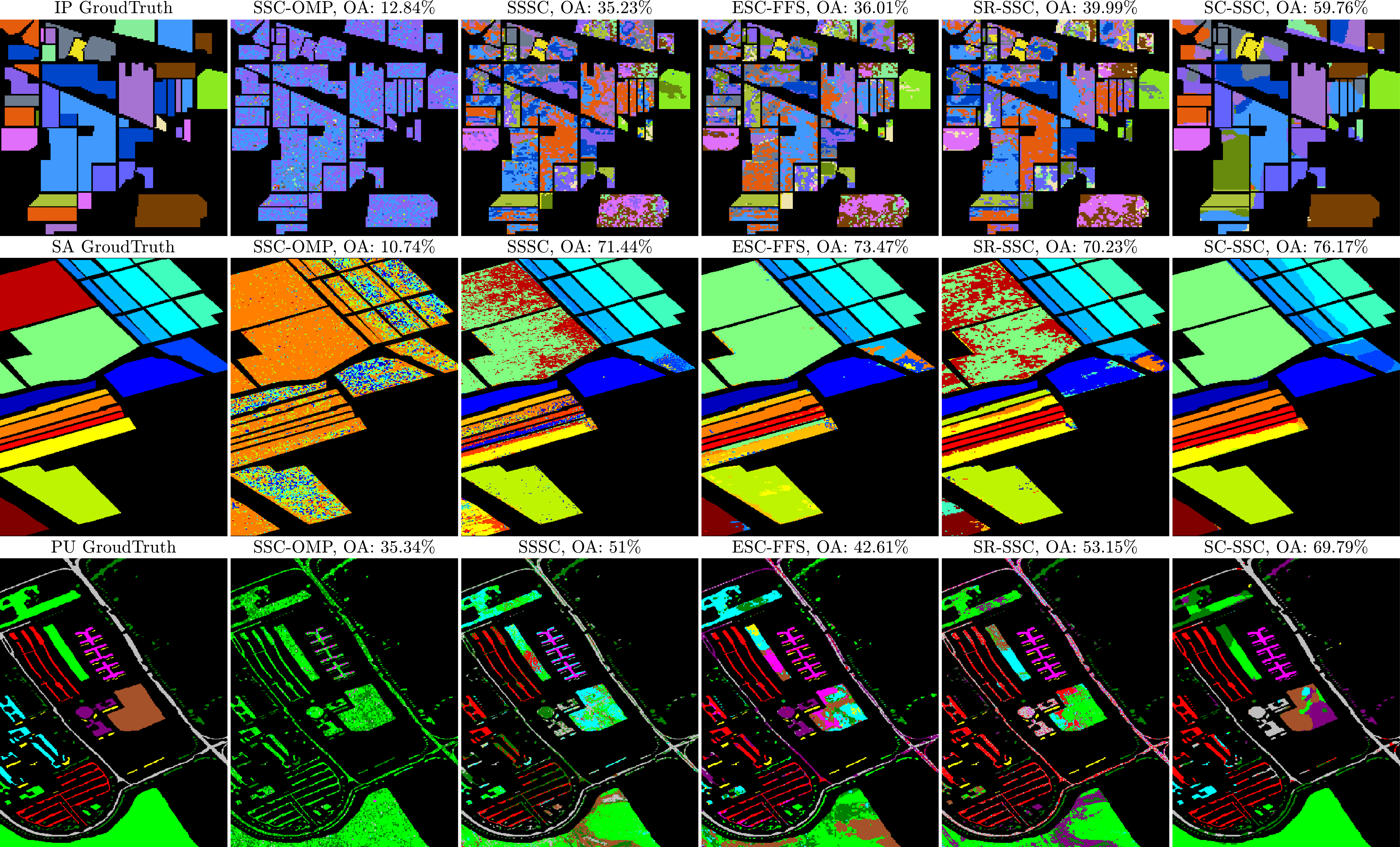}
	\caption{Land cover maps on the Indian Pines (IP), Salinas Valley (SA), and University of Pavia (PU) Full images. The proposed method is compared only with the scalable SSC-based methods we found in the literature.}
	\label{fig:visual_results_full}
\end{figure*}

We now compare the performance of SC-SSC with the scalable approaches: SSC-OMP, SSSC, ESC-FFS, and SR-SSC. Figure \ref{fig:visual_results_full} and Table \ref{table:QuantRes_full} present the visual and quantitative results respectively on the full spectral images. From both, qualitative and quantitative results, we observed that the proposed SC-SSC method outperforms the other approaches in terms of OA, Kappa and NMI score. Note that, although the proposed method is not the fastest one, it provides high clustering performance in a shorter amount of time in comparison with other methods.

\begin{table}[t]
	\centering
	{\footnotesize \caption{Quantitative results of Indian Pines, Salinas, and University of Pavia ROIs.}
	\label{table:QuantRes}}
	\resizebox{0.8\columnwidth}{!}{%
		\setlength{\tabcolsep}{3pt}
		\def\arraystretch{0.95}%
		\begin{tabular}{@{}cccccc@{}}
			\toprule
			\textbf{Class}    & \textbf{SSC}    & \textbf{S-SSC}  & \textbf{ORGEN}  & \textbf{3DS-SSC} & \textbf{SC-SSC} \\ \midrule
			\multicolumn{6}{c}{Indian Pines ROI}                                                                         \\ \midrule
			Corn-no-till      & \textbf{97.43}  & 83.98           & 93.19           & 73.24            & {\ul 95.53}     \\
			Grass             & 89.14           & {\ul 89.50}     & \textbf{92.47}  & 83.64            & 76.50           \\
			Soybean-no-till   & 41.36           & 62.38           & 54.33           & {\ul 73.73}      & \textbf{98.10}  \\
			Soybeans-min-till & 61.57           & 75.93           & 64.34           & {\ul 86.62}      & \textbf{94.47}  \\ \midrule
			AA                & 69.35           & 79.09           & 71.84           & {\ul 82.07}      & \textbf{94.69}  \\
			OA                & 62.62           & 76.16           & 69.91           & {\ul 80.41}      & \textbf{93.14}  \\
			Kappa             & 0.48            & 0.66            & 0.56            & {\ul 0.72}       & \textbf{0.90}   \\
			NMI               & 0.39            & 0.47            & 0.42            & {\ul 0.57}       & \textbf{0.79}   \\ \midrule
			Time {[}s{]}      & {\ul 285.57}    & 301.41          & 668.36          & 341.21           & \textbf{1.63}   \\ \midrule
			\multicolumn{6}{c}{Salinas ROI}                                                                              \\ \midrule
			Brocoli-1         & \textbf{100.00} & \textbf{100.00} & \textbf{100.00} & {\ul 87.28}      & \textbf{100.00} \\
			Corn-senesced     & \textbf{100.00} & 99.45           & {\ul 99.63}     & 99.46            & \textbf{100.00} \\
			Lettuce-4wk       & 0.00            & 44.79           & 45.86           & {\ul 53.24}      & \textbf{97.88}  \\
			Lettuce-5wk       & 70.19           & 97.72           & 98.74           & \textbf{100.00}  & {\ul 98.87}     \\
			Lettuce-6wk       & {\ul 98.48}     & 84.86           & 96.58           & \textbf{100.00}  & \textbf{100.00} \\
			Lettuce-7wk       & 98.97           & {\ul 99.48}     & 99.07           & 99.16            & \textbf{100.00} \\ \midrule
			AA                & 75.75           & 88.13           & 89.51           & {\ul 91.42}      & \textbf{99.37}  \\
			OA                & 77.16           & 83.09           & 85.27           & {\ul 88.26}      & \textbf{99.42}  \\
			Kappa             & 0.71            & 0.79            & 0.82            & {\ul 0.86}       & \textbf{0.99}   \\
			NMI               & 0.85            & 0.84            & {\ul 0.87}      & {\ul 0.87}       & \textbf{0.98}   \\ \midrule
			Time {[}s{]}      & {\ul 319.42}    & 327.66          & 1355.79         & 377.11           & \textbf{2.06}   \\ \midrule
			\multicolumn{6}{c}{University of Pavia ROI}                                                                  \\ \midrule
			Asphalt           & 1.47            & 53.90           & 33.38           & \textbf{93.56}   & {\ul 92.17}     \\
			Meadows           & 30.02           & 37.34           & {\ul 49.64}     & 30.87            & \textbf{93.87}  \\
			Gravel            & \textbf{6.48}   & {\ul 0.61}      & 0.00            & 0.00             & 0.00            \\
			Trees             & 90.37           & 0.00            & 97.08           & \textbf{100.00}  & {\ul 97.92}     \\
			Metal Sheets      & 18.48           & 79.03           & \textbf{100.00} & 86.21            & {\ul 98.74}     \\
			Bare Soil         & 95.53           & 89.28           & \textbf{99.71}  & 93.94            & {\ul 98.75}     \\
			Bitumen           & 47.15           & {\ul 72.13}     & 64.55           & 50.42            & \textbf{83.60}  \\
			Bricks            & 0.00            & {\ul 51.69}     & 0.06            & 0.00             & \textbf{72.36}  \\
			Shadows           & 22.77           & 0.00            & \textbf{100.00} & {\ul 53.27}      & 28.57           \\ \midrule
			AA                & 42.71           & 43.26           & {\ul 63.17}     & 62.01            & \textbf{72.98}  \\
			OA                & 32.68           & 46.59           & 58.49           & {\ul 65.07}      & \textbf{77.57}  \\
			Kappa             & 0.23            & 0.38            & 0.50            & {\ul 0.56}       & \textbf{0.72}   \\
			NMI               & 0.41            & 0.49            & 0.59            & {\ul 0.65}       & \textbf{0.82}   \\ \midrule
			Time {[}s{]}      & 17821.25        & 10195.89        & {\ul 551.02}    & 10501.29         & \textbf{68.72}  \\ \bottomrule
	\end{tabular}}
\end{table}

\begin{table}[h]
	\centering
	{\footnotesize
		\caption{Quantitative comparison with unsupervised deep learning-based methods in terms of NMI score.}
		\label{tab:deep-cmp}}
	\resizebox{\columnwidth}{!}{%
		\setlength{\tabcolsep}{3.5pt}
		\begin{tabular}{c|ccccc}
			\hline
			\backslashbox{Datasets}{Methods}       & VAE \cite{tulczyjew2020unsupervised}   & 3D-CAE \cite{nalepa2020unsupervised} & AE-GRU \cite{tulczyjew2020unsupervised} & AE-LSTM \cite{tulczyjew2020unsupervised} & SC-SSC \\ \hline
			Indian Pines  & 0.429 &  0.504  & {\ul 0.515}  & 0.478 & \textbf{0.601}   \\ \hline
			Salinas      & 0.722 & {\ul 0.839}  & 0.825  & 0.830   & \textbf{0.892}   \\ \hline
			University of Pavia   & 0.505  & {\ul 0.639}     & 0.524    & 0.569     & \textbf{0.643} \\ \hline
	\end{tabular}}
\end{table}

\begin{table}[t]
	\centering
	{\footnotesize \caption{Quantitative results of Indian Pines, Salinas, and University of Pavia Full Images.} \label{table:QuantRes_full}}
	\resizebox{0.9\columnwidth}{!}{%
		\setlength{\tabcolsep}{3pt}
		\def\arraystretch{0.95}%
	\begin{tabular}{@{}cccccc@{}}
		\toprule
		\textbf{Class}              & \textbf{SSC-OMP} & \textbf{SSSC}  & \textbf{ESC-FFS} & \textbf{SR-SSC} & \textbf{SC-SSC} \\ \midrule
		\multicolumn{6}{c}{Indian Pines}                                                                                       \\ \midrule
		Alfalfa                     & \textbf{7.69}    & {\ul 6.90}     & 0.60             & 0.00            & 0.00            \\
		Corn-no-till                & 37.04            & 34.33          & {\ul 57.79}      & 47.95           & \textbf{66.96}  \\
		Corn-min-till               & {\ul 20.83}      & 17.46          & 17.45            & 17.84           & \textbf{55.25}  \\
		Corn                        & {\ul 22.73}      & 15.56          & 11.34            & 8.85            & \textbf{27.25}  \\
		Grass-pasture               & 21.05            & {\ul 35.53}    & 34.81            & 32.39           & \textbf{90.52}  \\
		Grass-trees                 & 57.14            & \textbf{84.91} & 77.94            & 70.11           & {\ul 77.05}     \\
		Grass-pasture-mowed         & \textbf{0.45}    & 0.00           & {\ul 0.29}       & 0.00            & 0.00            \\
		Hay-windrowed               & 25.81            & {\ul 88.29}    & 85.91            & 76.91           & \textbf{90.53}  \\
		Oats                        & 0.00             & {\ul 3.36}     & 0.02             & \textbf{4.36}   & 0.00            \\
		Soybean-no-till             & 18.18            & 30.89          & {\ul 38.54}      & 36.11           & \textbf{64.15}  \\
		Soybean-min-till            & 28.67            & 52.25          & {\ul 55.12}      & 48.52           & \textbf{62.23}  \\
		Soybean-clean               & 6.32             & {\ul 22.17}    & 17.44            & 15.76           & \textbf{32.10}  \\
		Wheat                       & 8.24             & 37.60          & 37.53            & {\ul 56.35}     & \textbf{66.13}  \\
		Woods                       & 11.54            & 88.22          & 81.03            & {\ul 89.42}     & \textbf{91.49}  \\
		Building-grass-trees-drives & 5.56             & 17.33          & {\ul 25.51}      & 22.15           & \textbf{67.34}  \\
		Stone-stell-towers          & 0.00             & {\ul 45.20}    & 18.53            & 38.14           & \textbf{49.73}  \\ \midrule
		AA                          & 9.64             & {\ul 40.55}    & 33.94            & 40.26           & \textbf{56.37}  \\
		OA                          & 12.84            & 35.23          & 36.01            & {\ul 39.99}     & \textbf{59.76}  \\
		Kappa                       & 0.03             & 0.29           & 0.30             & {\ul 0.33}      & \textbf{0.55}   \\
		NMI                         & 0.03             & 0.42           & 0.43             & {\ul 0.44}      & \textbf{0.60}   \\ \midrule
		Time {[}s{]}                & 37.93            & 32.35          & 47.15            & \textbf{16.39}  & {\ul 19.33}     \\ \midrule
		\multicolumn{6}{c}{Salinas}                                                                                            \\ \midrule
		Brocoli 1                   & 11.34            & 0.00           & \textbf{100.00}  & 0.00            & {\ul 98.38}     \\
		Brocoli 2                   & 28.88            & 57.41          & \textbf{99.68}   & {\ul 62.98}     & \textbf{99.68}  \\
		Fallow                      & 9.70             & {\ul 78.98}    & 72.16            & 0.00            & \textbf{99.88}  \\
		Fallow Plow                 & 7.43             & {\ul 91.56}    & 89.52            & \textbf{94.35}  & 47.71           \\
		Fallow Smooth               & 22.58            & 58.11          & \textbf{71.98}   & {\ul 71.82}     & 64.94           \\
		Stubble                     & 21.54            & 99.05          & {\ul 99.70}      & \textbf{99.91}  & 95.67           \\
		Celery                      & 22.08            & 97.12          & {\ul 97.78}      & 85.62           & \textbf{99.94}  \\
		Grapes                      & 3.47             & {\ul 70.19}    & 58.44            & \textbf{70.68}  & 60.92           \\
		Soil                        & 24.68            & 91.18          & {\ul 95.91}      & 85.58           & \textbf{96.88}  \\
		Corn                        & 3.91             & {\ul 61.92}    & 26.70            & 58.04           & \textbf{99.76}  \\
		Lettuce 4                   & 3.97             & \textbf{79.10} & {\ul 24.01}      & 0.00            & 0.00            \\
		Lettuce 5                   & 4.42             & \textbf{81.09} & {\ul 65.14}      & 51.99           & 62.81           \\
		Lettuce 6                   & {\ul 1.90}       & \textbf{41.17} & -                & 0.00            & 0.00            \\
		Lettuce 7                   & 0.00             & \textbf{58.92} & {\ul 55.09}      & 51.18           & 43.70           \\
		Vineyard                    & 11.57            & \textbf{56.92} & 25.00            & {\ul 49.22}     & 0.00            \\
		Vineyard trellis            & 3.05             & 0.00           & 98.53            & {\ul 98.89}     & \textbf{100.00} \\ \midrule
		AA                          & 10.95            & 64.76          & \textbf{73.15}   & 61.57           & {\ul 73.03}     \\
		OA                          & 10.74            & 71.44          & {\ul 73.47}      & 70.23           & \textbf{76.17}  \\
		Kappa                       & 0.05             & 0.68           & {\ul 0.70}       & 0.67            & \textbf{0.73}   \\
		NMI                         & 0.16             & 0.75           & {\ul 0.83}       & 0.78            & \textbf{0.87}   \\ \midrule
		Time {[}s{]}                & \textbf{78.18}   & {\ul 58.29}    & 615.40           & 168.18          & 906.71          \\ \midrule
		\multicolumn{6}{c}{University of Pavia}                                                                                \\ \midrule
		Asphalt                     & 20.67            & 60.00          & \textbf{95.60}   & 68.64           & {\ul 73.52}     \\
		Meadows                     & 48.11            & {\ul 85.58}    & 71.91            & 77.98           & \textbf{96.81}  \\
		Gravel                      & \textbf{13.73}   & 0.15           & 0.06             & 1.25            & {\ul 11.73}     \\
		Trees                       & 9.54             & 31.28          & 18.03            & \textbf{72.63}  & {\ul 32.48}     \\
		Metal sheets                & 81.71            & \textbf{97.01} & 36.74            & 60.15           & {\ul 86.22}     \\
		Bare soil                   & 0.00             & 5.75           & {\ul 24.15}      & 15.04           & \textbf{97.53}  \\
		Bitumen                     & 0.00             & {\ul 4.26}     & \textbf{27.80}   & 0.00            & 0.00            \\
		Bricks                      & 0.00             & {\ul 55.27}    & \textbf{60.70}   & 39.11           & 54.54           \\
		Shadows                     & 0.00             & 43.32          & {\ul 62.99}      & \textbf{96.45}  & 0.00            \\ \midrule
		AA                          & 16.48            & 39.76          & \textbf{56.87}   & {\ul 53.50}     & 52.23           \\
		OA                          & 35.34            & 51.00          & 42.61            & {\ul 53.15}     & \textbf{69.79}  \\
		Kappa                       & 0.07             & 0.39           & 0.33             & {\ul 0.42}      & \textbf{0.61}   \\
		NMI                         & 0.07             & 0.39           & 0.49             & {\ul 0.52}      & \textbf{0.64}   \\ \midrule
		Time {[}s{]}                & 201.52           & \textbf{73.45} & 1821.58          & {\ul 148.55}    & 913.67          \\ \bottomrule
	\end{tabular}
	}
\vspace{-0.1in}
\end{table}

\subsubsection{Comparison with unsupervised deep-learning-based methods}
For the sake of completeness, we compare the proposed SC-SSC method with unsupervised deep-learning-based methods based on autoencoders (AE) for spectral image clustering. Three of them were proposed in \cite{tulczyjew2020unsupervised} (VAE, AE-GRU, and AE-LSTM), and the 3D-CAE method was proposed in \cite{nalepa2020unsupervised} which is based on a 3D convolutional AE. Note that we only compare our method with totally unsupervised deep learning approaches to make a fair comparison. Table \ref{tab:deep-cmp} shows the quantitative results in terms of the NMI score. In the table, the best result is shown in bold font, and the second-best is underlined. As observed, our method obtains an NMI score of $0.601$, $0.892$, and $0.643$ on Indian Pines, Salinas, and University of Pavia full spectral images, respectively, corresponding to the highest clustering scores.

\section{Conclusion}
\label{sec:conclusions}
In this work, we presented a new subspace clustering algorithm for land cover segmentation which can handle large-scale datasets and take advantage of spectral images' neighboring spatial information to boost the clustering accuracy. Our method considers the spatial similarity among spectral pixels to select the most representative ones, such that all other neighboring points can be well-represented by those representative pixels in terms of a sparse representation cost. Then, the obtained sparse coefficients matrix is enhanced by performing filtering on the coefficients, and a fast spectral clustering algorithm gives the segmentation. Through simulations using traditional test spectral images, we demonstrated the effectiveness of our method for fast land cover segmentation, obtaining remarkable high clustering performance when compared with state-of-the-art SSC algorithms and even novel unsupervised-deep-learning-based methods.

\vspace{-0.2in}

%




%
%

\ifCLASSOPTIONcaptionsoff
  \newpage
\fi



\bibliographystyle{IEEEtran}
\bibliography{biblio}
\end{document}